\documentstyle[prd,aps,epsfig,preprint,tighten]{revtex}

\begin{document}

\preprint{                                                 BARI-TH/335-99}
\draft
\title{  Testing violations of special and general relativity\\
through the energy dependence of $\nu_\mu\leftrightarrow\nu_\tau$ oscillations\\
in the Super-Kamiokande atmospheric neutrino experiment}

\author{     G.~L.~Fogli, E.~Lisi, A.~Marrone, and G.~Scioscia}
\address{   Dipartimento di Fisica and Sezione INFN di Bari\\ 
                  Via Amendola 173, I-70126 Bari, Italy}
\maketitle
\begin{abstract}
The atmospheric neutrino data collected by the Super-Kamiokande experiment span
about four decades in neutrino energy $E$,  and are thus appropriate to probe
the energy dependence of the   oscillation wavelength $\lambda$ associated to
$\nu_\mu\leftrightarrow\nu_\tau$ flavor transitions, when these are assumed to
explain the data. Such dependence takes the form  $\lambda^{-1}\propto E^n$ in
a wide class of theoretical models, including ``standard'' oscillations due to
neutrino mass and mixing $(n=-1)$, energy-independent oscillations $(n=0)$, and
violations of the equivalence  principle or of Lorentz invariance $(n=1)$. We
study first how the theoretical zenith distributions of sub-GeV, multi-GeV, and
upward-going muon events change for different integer values of $n$. Then we
perform a detailed analysis of the  Super-Kamiokande data by treating the
energy exponent $n$ as a  free parameter, with unconstrained scale factors for
both the amplitude and the phase of $\nu_\mu\leftrightarrow\nu_\tau$
oscillations. We find a best-fit range $n=-0.9\pm0.4$ at 90\% C.L., which
confirms the standard  scenario $(n=-1)$ as the dominant oscillation 
mechanism, and strongly constrains possible concurrent exotic processes ($n\neq
-1$). In particular, we work out the interesting case of leading standard
oscillations  plus subleading terms induced by violations of special or
general  relativity principles, and obtain  extremely stringent upper bounds on
the amplitude of such violations in the $(\nu_\mu,\nu_\tau)$ sector. 
\end{abstract}

\section{Introduction}

The recent atmospheric neutrino data from the Super-Kamiokande (SK) experiment
\cite{To98} can be beautifully explained through flavor oscillations generated
by nonzero neutrino mass and mixing \cite{Po67,Ka62} in the
$\nu_\mu\leftrightarrow\nu_\tau$ channel \cite{SKEV}. Such interpretation is
consistent with all the SK data, including sub-GeV  $e$-like and $\mu$-like
events (SG$e,\mu$) \cite{SKSG}, multi-GeV  $e$-like and $\mu$-like events
(MG$e,\mu$) \cite{SKMG}, and upward-going muon events (UP$\mu$) \cite{SKUP}. A
combined analysis of the 33 kTy SK data sample can be found in \cite{Fo99}. The
oscillation hypothesis has been strengthened by the latest (preliminary) 45 kTy
SK data sample \cite{Me99,Ha99}, and is also consistent with independent
atmospheric neutrino results from the MACRO \cite{MACR} and Soudan-2
\cite{SOUD} experiments, as well as with the finalized upward-going muon data
from the pioneering Kamiokande experiment \cite{KAUP}.

Establishing $\nu_\mu\leftrightarrow\nu_\tau$ oscillations generated by nonzero
$\nu$ mass and mixing as the ``standard'' interpretation requires, however,
further data and analyses. Basically, the following three aspects  should be
clarified: (1) the periodicity; (2) the flavors; and (3)  the dynamics.

So far, the periodicity of the $\nu_\mu$ oscillation pattern has not been
experimentally observed in the neutrino energy $(E)$ or pathlength $(L)$
domain, and it is unlikely to emerge from the SK lepton distributions,  largely
smeared in energy or angle. Although specific nonperiodic scenarios, such as
neutrino decay \cite{Ba99}, can be indirectly excluded by careful analyses of
SK data \cite{Li99,Sc99}, the direct observation of a periodic disappearance
pattern of $\nu_\mu$'s remains an important goal for future atmospheric and
long-baseline neutrino experiments.

Besides periodicity, one should also identify unambiguously the flavor(s) of
the oscillating partner(s) of $\nu_\mu$'s because, in principle, all
oscillation channels into active or sterile neutrinos might be open
($\nu_\mu\leftrightarrow\nu_\tau$, $\nu_\mu\leftrightarrow\nu_e$,
$\nu_\mu\leftrightarrow\nu_s$). While the amplitude of possible 
$\nu_\mu\leftrightarrow\nu_e$ transitions is bound to be small \cite{Fo99}, one
cannot exclude $\nu_\mu\leftrightarrow\nu_s$ oscillations with large amplitude
with present SK data \cite{Me99,Ha99}. However, based on the fact that
different oscillation channels induce somewhat different energy-angle lepton
distributions, there are good prospects for a  significant ($>3\sigma$)
discrimination of $\nu_\mu\leftrightarrow\nu_\tau$ from
$\nu_\mu\leftrightarrow\nu_s$ with future SK data \cite{Ka99}.

In this work we assume that both the periodicity (i.e., the existence of an
oscillation length $\lambda$) and the oscillating flavors ($\nu_\mu,\nu_\tau$)
are established, and we rather focus on the third aspect to be clarified: The
dynamical origin of $\nu_\mu\leftrightarrow\nu_\tau$ oscillations. The
standard  oscillation dynamics, involving a nontrivial $2\times2$ neutrino mass
matrix, leads to a well-known energy dependence of $\lambda$,
\begin{equation}
\lambda^{-1}\propto E^{-1}\ {\rm (standard)}\ .
\label{std}
\end{equation}
However, possible nonstandard neutrino interactions or properties can also
generate (or coexist with) $\nu_\mu\leftrightarrow\nu_\tau$ oscillations 
\cite{Gr95,Jo98}. An incomplete list of possibilities include violations of the
equivalence principle (VEP) \cite{Ga88,Ga89}, flavor-changing neutral currents
(FCNC) (see \cite{Go99} and refs.\ therein), neutrino couplings to  space-time
torsion fields \cite{DS81}, neutrino interactions through charged scalar
particles \cite{Fu88}, nonrelativistic heavy neutrinos \cite{Ah97}, violations
of Lorentz invariance (VLI) \cite{Co97,Gl97} and of CPT symmetry \cite{Co98}.
In several such models the energy dependence of $\lambda$ takes a power-law
form \cite{Ya94},
\begin{equation}
\lambda^{-1}\propto E^{n}\ (n\neq -1,{\rm\ nonstandard)}\ .
\label{nonstd}
\end{equation}

Although models with exotic dynamics for $\nu_\mu\leftrightarrow\nu_\tau$ 
oscillations do not survive Occam's razor, they might survive experimental
tests! Effective tests must cover the widest possible energy range, as  evident
from Eqs.~(\ref{std}) and (\ref{nonstd}). Concerning atmospheric neutrinos,
pre-SK data analyses covered only about two decades in energy (i.e., the
so-called contained events, $E\sim 0.1$--$10$ GeV), and were compatible with
several nonstandard scenarios, in particular with the VEP hypothesis
(corresponding to $n=1$) \cite{Pa93,Ha96}. An interesting post-SK analysis,
covering a slightly more extended energy range \cite{Fo98}, does not appear to
discriminate significantly the three cases $n=0$ and $n=\pm1$ examined. 
However, as observed in \cite{Li99,Sc99}, a much longer ``lever arm'' in the
energy domain is provided by the inclusion of partially contained and
upward-going muon events (up to $E\sim 10^3$ GeV), thus providing a powerful
tool to test exotic scenarios (which, indeed, appear to be  disfavored in
general \cite{Li99}).

In this work we assess quantitatively the situation for 
$\nu_\mu\leftrightarrow\nu_\tau$ models with power-law energy dependence of the
oscillation length  ($\lambda^{-1}\propto E^n$), including, as relevant
subcases, standard mass-mixing oscillations $(n=-1)$ and violations of the
equivalence principle or of Lorentz invariance $(n=1)$.  We obtain two basic
results: (1) The 90\% C.L. range of $n$ is determined to be $n=-0.9\pm0.4$,
which is in striking agreement with standard oscillations, and excludes all
$n\neq -1$ models as dominant sources of $\nu_\mu\leftrightarrow\nu_\tau$
transitions; (2) assuming the $n=1$ case as a subleading mechanism (coexisting
with leading standard oscillations), we place stringent upper bounds on its
amplitude. Such bounds can be interpreted as upper limits to violations of
special or general relativity principles.

The plan of the paper is as follows. In Sec.~II we analyze models with generic 
$\lambda^{-1}\propto E^n$ behavior, and  constrain $n$ through fits to SK
atmospheric $\nu$ data. In Sec.~III we consider a more complicated case with
two coexisting sources for oscillations, namely, neutrino masses and violations
of relativity principles. We summarize our results in Sec.~IV.

\section{Analysis of models with $\lambda^{-1}\propto E^{\lowercase{n}}$}

In this section, we first review some oscillation models with $\lambda$
depending on $E$ through a power law. Then, independently of specific models,
we study the phenomenology of atmospheric $\nu$'s, and constrain the energy
exponent $n$ through detailed fits to the SK data.

\subsection{Review of models}

Typical two-flavor hamiltonians for  $\nu_\mu\leftrightarrow\nu_\tau$
oscillations predict a flavor transition probability of the form
\begin{equation}
P(\nu_\mu\leftrightarrow\nu_\tau)=
\sin^2 2\xi \, \sin^2  (\pi \lambda^{-1}L)\ ,
\label{P}
\end{equation}
where $\xi$ is the rotation (mixing) angle between the flavor basis and the
basis where the hamiltonian is diagonal, $L$ is the neutrino  pathlength, and
$\lambda$ is the neutrino oscillation length. In several cases of interest,
$\lambda$ depends on the neutrino energy $E$ as
\cite{Ya94,Fo98}
\begin{equation}
\lambda^{-1} \propto E^n
\label{lambda}
\end{equation}
with the exponent $n$ taking integer (positive or negative) values.

For instance, the cases $n=-1$, 0 and 1 arise, respectively,  in the presence
of neutrino interactions mediated by scalar, vector, and tensor fields
\cite{Fo98}. Specific models include:
\begin{mathletters}
\begin{eqnarray}
\lambda^{-1}&=&\frac{\Delta m^2}{4\pi E}\ \ \ \ \ \ \ \ \ (n=-1,
{\rm\ standard\ oscillations\ }\protect\cite{Po67})\ ,
\label{a}\\
&=& \frac{E|\phi|\Delta\gamma}{\pi} \ \ \ \ \ (n=1,
{\rm\ violations\ of\ equivalence\ principle\ }\protect\cite{Ga88})\ ,
\label{b}\\
&=& \frac{E \delta v}{2\pi} \ \ \ \ \ \ \  \ \ \ (n=1,
{\rm\ violations\ of\ Lorentz\ invariance\ }\protect\cite{Co97})\ ,
\label{c}\\
&=& \frac{\delta b}{2\pi} \ \ \ \ \ \ \ \ \ \ \ \ (n=0,
{\rm\ violations\ of\ CPT\ symmetry\ }\protect\cite{Co98})\ ,
\label{d}\\
&=& \frac{Q\delta k}{2\pi} \ \ \ \ \ \ \ \ \ \ (n=0,
{\rm\ nonuniversal\ coupling\ to\ a\ torsion\ field\ }\protect\cite{DS81})\ .
\label{e}
\end{eqnarray}
\end{mathletters}

Standard oscillations [Eq.~(\ref{a})] are simply generated by nonzero neutrino
masses ($\Delta m^2=m^2_2-m^2_1$), with a mixing angle $\xi$ [Eq.~(\ref{P})]
usually denoted  as $\theta_{23}$, $\psi$, or simply $\theta$ (we adopt
the symbol  $\psi$ in the following, as we use $\theta$ for the lepton zenith
angle). In this case, the energy exponent is $n=-1$.

Equation (\ref{b}) refers to possible violations of the equivalence principle
(VEP) \cite{Ga88}, namely, to nonuniversal couplings of neutrinos   to the
gravitational potential  $\phi$. The difference in such couplings is usually
denoted by $\Delta\gamma$, and the mixing angle by $\theta_G$. The potential
seems to be dominated by the local supercluster ($\phi\sim 3\times 10^{-5}$
\cite{Ke90}), but ambiguities in its definition suggest to use the product
$\phi\Delta\gamma$ as a single, dimensionless free parameter, rather than
$\phi$ and $\Delta\gamma$ separately  (see \cite{Ha96,Ba95,Mu97} for recent
discussions). The energy exponent of VEP-induced oscillations is $n=1$.%
\footnote{An alternative,  string-inspired VEP mechanism \protect\cite{Da94},
leading to an energy exponent $n=-1$ rather than $n=1$, has been recently
considered in \protect\cite{Ha98}. Its phenomenology would be indistinguishable
from the standard case, as far as $\nu_\mu\leftrightarrow\nu_\tau$ oscillations
are involved. We do not consider such VEP scenario in this work.}

Equation (\ref{c}) refers to possible violations of Lorentz invariance (VLI),
namely, to asymptotic neutrino velocities different from $c$ \cite{Co97}. The
parameter $\delta v$ represents the speed difference in units of $c$. The
mixing angle is usually denoted as $\theta_v$. Since the energy exponent $n$ is
the same ($=1$) in both the VEP and the VLI scenario, such mechanisms are
phenomenologically equivalent through the substitutions $|\phi|\Delta\gamma \to
\delta v/2$ and $\theta_G\to\theta_v$ \cite{Gl97}.

Equation (\ref{d}) refers to possible violations of the CPT symmetry through a
more general class of Lorentz-violating perturbations \cite{Co98}, $\delta b$
being proportional to the CPT-odd hamiltonian producing energy independent
$(n=0)$ oscillations.  It is interesting to notice that the most general
Hamiltonian considered in \cite{Co98} encompasses the three scenarios with
$n=-1$, 0 and 1.

Finally, Eq.~(\ref{e}) refers to possible nonuniversal couplings $(\Delta k\neq
0)$ of neutrinos to a space-time torsion field of strength $Q$ \cite{DS81},
which also produce energy-independent oscillations.

In principle,  several oscillation mechanisms with different $n$'s  might occur
at the same time, with corresponding complications in the analysis. In the next
two subsections, however, we consider only  one mechanism at a time. We will
discuss the interesting case of  coexisting $n=-1$ plus $n=+1$ oscillations in
Sec.~III.

A remark is in order. Not all ``exotic'' models for
$\nu_\mu\leftrightarrow\nu_\tau$ transitions can be parametrized as in
Eqs.~(\ref{P}) and (\ref{lambda}). An important exception is represented by
FCNC-induced oscillations  (see \cite{Go99} and references therein).  In fact,
although the FCNC oscillation phase is energy-independent $(n=0)$, it is
proportional to the column density of electrons rather than to $L$. Therefore,
FCNC-induced oscillations deserve a separate analysis \cite{Go99,Li99} and will
be considered in a future work. However, some features of the  $n=0$ case also
apply qualitatively to the FCNC case.

Finally, we note in passing that the analysis of nonstandard energy dependences
for $\nu$ oscillations has some correspondence in the neutral kaon system,
where the role of $\lambda$ is played by the effective  ${\rm
K}{}^0$--$\overline{\rm K}{}^0$ mass difference (see \cite{El99} and references
therein).

\subsection{Model-independent analysis}

In this section we do not commit ourselves to any specific model, and rather
analyze the phenomenology of the $\lambda^{-1}\propto E^n$ dependence in the
most general way. We assume that the $\nu_\mu\leftrightarrow\nu_\tau$
oscillation probability takes the form
\begin{equation}
P(\nu_\mu\leftrightarrow\nu_\tau)=\alpha \cdot \sin^2
\left( \beta\;\frac{L}{10^3{\rm\ km}}\;\frac{E^n}{{\rm GeV}^n}\right)\ ,
\label{psk}
\end{equation}
where $\alpha$ is an overall scale factor ($0\leq \alpha\leq 1$) for the
oscillation amplitude, $\beta$ is an unconstrained scale factor for the 
oscillation phase, and $n$ is a free exponent  (not necessarily equal to $-1$,
0, or 1).  The units in Eq.~(\ref{psk}) have been conveniently chosen on the
basis of the current Super-Kamiokande data, which suggest an oscillation length
of $O(10^3{\rm\ km})$ for neutrino energies of $O(1 {\rm\ GeV})$.  The standard
oscillation case is recovered by taking $n=-1$, $\alpha=\sin^2 2\psi$, and
$\beta =1.27\, \Delta m^2/(10^{-3} {\rm\ eV}^2)$.

Equation~(\ref{psk}) is used to calculate the observable rates of sub-GeV,
multi-GeV, and upward through-going muon events in SK  as a function of the
lepton zenith angle $\theta$, using the same detailed and accurate approach as
in Refs.~\cite{Fo99,Sc99}, to which  we refer the reader for technical
details.  Here we just remind that the parent neutrino distributions for
sub-GeV (SG), multi-GeV  fully contained (MG~FC), multi-GeV partially contained
(MG~PC), and upward through-going (UP) muon events are  roughly distributed in
the ranges $0.1$--$3$ GeV, $1$--$10$ GeV, $1$--$10^2$ GeV, and $10$--$10^3$
GeV, respectively, thus covering about four decades in $E$.

Figure~1 shows  the expected zenith distributions of SG$\mu$, MG$\mu$ (FC+PC), 
and UP$\mu$ events in Super-Kamiokande for maximal mixing  ($\alpha=1$) and for
three representative values of $\beta$:  dotted line, $\beta=0.2$; dashed line,
$\beta=1$;  solid line, $\beta=5$ (on color printers such lines appear in
green, red, and blue, respectively). In each bin, the muon rates $\mu$  are
normalized to the expectations $\mu_0$ in the absence of oscillations, so that
deviations from the no-oscillation case ($\mu/\mu_0=1$) are immediately
recognizable. Since we are considering pure $\nu_\mu\leftrightarrow\nu_\tau$
oscillations ($e/e_0=1$ always), the electron rates are not shown.  The effect
of taking $\alpha<1$ can be approximately described by reducing proportionally
the amount of muon disappearance in each subfigure.  The theoretical
expectations are affected by relatively large and strongly correlated
uncertainties (not shown), mainly related to the overall normalization of the
atmospheric neutrino flux \cite{Fo99}. The 45 kTy SK data \cite{Me99,Ha99}
(dots with statistical error bars) are superposed to guide the eye, although no
fit to the data is implied by this figure.

In Fig.~1, from top to bottom, the energy exponent $n$ takes the values $n=-2$,
$-1$, 0, and 1, corresponding to a dependence of the kind $L/E^2$, $L/E$, $L$,
and $L\cdot E$ for the oscillation phase. Such four cases are described in the
following.

{\em Case $L/E^2$.}  This  case ($n=-2$) does not correspond to any known
model, and is used only to extend the study to a power-law energy dependence
faster than in the standard case ($n=-1$). The muons appear to be significantly
suppressed at the lowest (SG) energies, the more the larger $\beta$. An
excessive suppression of SG muons is avoided only by keeping  $\beta \lesssim
1$. On the other hand, such values of $\beta$ are too low to produce a
significant suppression of MG muons, which rather prefer $\beta \gtrsim 5$.
Therefore, one expects a ``compromise'' between underestimated SG$\mu$ rate and
overestimated MG$\mu$ rates for $\beta$ in the range $\sim 1$--$5$. UP$\mu$
events are basically unsuppressed, due to their high energy.  Reducing $\alpha$
would be of no help in reducing the conflict between SG and MG muon
expectations.

{\em Case $L/E$.} This is the ``standard'' oscillation case, which,  as well
known, provides and excellent description of the SK data for $\Delta m^2\sim
{\rm\ few}\times 10^{-3}$ eV$^2$ \cite{SKEV,Fo99}.   Therefore, it is not
surprising to see in Fig.~1 that, for values of $\beta$ in the range $\sim
1$--$5$, all the muon data are well reproduced. Higher values of $\beta$ would
suppress SG muons too much, while lower values would not produce enough
suppression at higher energies (MG$\mu$ and UP$\mu$ samples).

{\em Case $L$.}  In this case, the $\nu_\mu$ survival probability is energy
independent, and the differences in the muon suppression patterns among the SG,
MG and UP muon samples are mainly due to the  different angular smearing. At
low (SG) energies the smearing is very effective and there is little difference
between the three curves, while higher-energy, MG muons are more
discriminating. Values of $\beta$ around unity (dashed curve)  seem to be
preferred by SG+MG data, and produce a significant $(\sim 1/2)$ suppression
below the horizon. UP muons have a different suppression pattern, which is
highly correlated with the parent neutrino direction, and follows closely the
variations of $P_{\mu\mu}$ with $L$ (dashed and dotted curves), until the
oscillations are so fast to be unresolved within the bin width (solid curve).
The expected UP$\mu$ suppression appears to be larger than suggested by the
data, unless $\alpha$ is taken to be nonmaximal; however, for $\alpha<1$ the
relatively good description of SG+MG data would be spoiled (not shown).

{\em Case $L\cdot E$.}  In this case, the $\nu_\mu$ survival probability
rapidly approaches the average value $1/2$ as the energy increases. Therefore,
although SG+MG data can be described relatively well with $\alpha=1$ and
$\beta\sim 0.2$, the expected UP$\mu$ rates  are too low in any case.  As in
the previous $n=0$ scenario, taking $\alpha<1$ would help to reproduce the
UP$\mu$ data, but would worsen the description of SG+MG data. No satisfactory
compromise can be reached.

In all the four cases, it can be seen that SG (or even SG+MG) data alone do not
discriminate strongly among the various scenarios, since they do not probe the
full energy range explorable by SK. This might explain why the SG+MG analysis
in \cite{Fo98} could not significantly distinguish the three cases $n=-1$, 0,
and 1. The inclusion of UP$\mu$ data extends the range up to $E\sim 10^3$ GeV,
and provides a very important tool to probe the energy dependence of $\lambda$
\cite{Li99}.

In conclusion, from the examination and the comparison of the oscillated muon
distributions at different values of $n$ shown in Fig.~1,  the case $n=-1$
emerges as a good description of the SK data at all energies, while for $n\neq
-1$ the patterns of  muon suppression at low (SG), intermediate
(MG) and high (UP) energies appear to be in
conflict with the data.

\subsection{Fits to the Super-Kamiokande data}

The qualitative understanding of the muon distributions at different values of
$n$ (previous subsection) can be improved by performing quantitative fits to
the 45 kTy preliminary SK data \cite{Me99,Ha99}.  We use a $\chi^2$ approach
that, as described in \cite{Fo99}, takes into account several sources of
correlation among the systematics affecting the theoretical predictions. Even
if $\nu_e$'s do not participate to oscillation in the scenarios considered
here, we have included  the SG and MG $e$-like data in the analysis (for a
total of 30 data points), since they play an important role in constraining the
overall normalization uncertainty (see also \cite{Fo99,Sc99}).

Figure~2 shows the best-fit zenith distributions of muons for the four cases
considered in Fig.~1 ($n=-2$, $-1$, 0, and 1), as obtained by leaving $\alpha$
and $\beta$ unconstrained. The values of $\alpha$, $\beta$, and $\chi^2$ at the
best-fit points are  reported in the top part of the  figure. The $L/E$
($n=-1$) case provides an excellent fit to the data ($\chi^2_{\rm min}/N_{\rm
DF}=20.3/28$), while all the other cases do not provide a good description of
at least one data sample (SG$\mu$, MG$\mu$, or UP$\mu$). In particular, for
$n=-2$ there is insufficient up-down asymmetry of MG$\mu$'s and no slope of
UP$\mu$'s; for $n=0$ none of the zenith distributions is correctly reproduced;
for $n=1$ there is a too strong and flat suppression of UP$\mu$'s.

In Fig.~3 we present the $\chi^2$ curve as obtained by taking also $n$ as a
free parameter, besides $\alpha$ and $\beta$. Although noninteger values of $n$
may not be related to any realistic oscillation dynamics, this exercise is
useful to see how accurately $n$ is determined through the SK data. The result
is striking:
\begin{equation}
n=-0.9\pm0.4 {\rm\ \ at\ 90\%\ C.L.} 
\label{n}
\end{equation}
(corresponding to $\chi^2-\chi^2_{\rm min }= 6.25$ for $N_{\rm DF}=3$). This
narrow range for $n$ is perfectly consistent with standard ($n=-1$)  neutrino
oscillations, and inconsistent with any other integer value of $n$.

Given the importance of standard $\nu_\mu\leftrightarrow\nu_\tau$ oscillations,
we show in Fig.~4 the updated limits on the oscillation parameters $\Delta m^2$
and $\sin^2 2\psi$. We find the best fit at $\Delta m^2=2.8\times 10^{-3}$
eV$^2$ and maximal mixing. The bounds in Fig.~4 are in good agreement with the
latest full data  analysis from the SK collaboration \cite{Ha99}.

In conclusion, standard oscillations $(\lambda^{-1}\propto E^{-1})$ are
strongly favored as the {\em dominant\/} mechanism for the 
$\nu_\mu\leftrightarrow\nu_\tau$ flavor transitions of atmospheric neutrinos in
SK. Alternative mechanisms of the kind $\lambda^{-1}\propto E^{n}$ (with
$n\neq -1)$ {\em cannot\/} be the dominant source of muon disappearance in SK.
In particular, violations of special or general relativity principles (VLI or
VPE, leading to  $\lambda^{-1} \propto E$)  cannot explain the bulk of SK
atmospheric $\nu$ data. Therefore, if $n\neq -1$ oscillations occur in nature,
they can only be subleading processes with small amplitude, coexisting with 
leading, large-amplitude $n=-1$ standard oscillations. Such results
generalize and refine previous indications that SK
data could disfavor some exotic models \cite{Li99}.

\section{Constraints on  violations of relativity principles}

In this section we consider a more complicated case, characterized by leading
$n=-1$ oscillations plus subleading $n=+1$ oscillations, possibly generated by
violations of relativity principles (VEP or VLI).   We show that the fit to the SK data is not improved with
respect to the case of standard  $\nu_\mu\leftrightarrow\nu_\tau$ transitions.
As a consequence, we derive upper bounds on such violations. A brief review of
the theoretical formalism precedes the phenomenological analysis.

\subsection{Violations of relativity principles: Formalisms}

The theory and phenomenology of neutrino violations of the equivalence
principle (VEP)  \cite{Ga88} have been investigated in a number of papers,
including studies of the   solar $\nu$ deficit 
\cite{Pa93,Ha96,Ba95,Mu97,Ha91,Bu93,Mi95,Mu96,Ma98,Ca99}, of the atmospheric 
$\nu$ anomaly \cite{Ya94,Pa93,Ha96,Fo98}, of oscillation searches at short
baseline \cite{Ma96} and long baseline  \cite{Ya94,Ha96,Ii93} accelerator
facilities, and of double beta decay \cite{Kl99}. Given the phenomenological
equivalence of violations of Lorentz invariance (VLI) \cite{Co97} and of the
equivalence principle (VEP) \cite{Gl97}, neutrino oscillation searches can be
generally interpreted as tests of fundamental principles of both special and
general relativity, with a sensitivity at levels below $10^{-20}$  (see, e.g.,
\cite{Co98,Ha96}). Here we focus on  the case of VEP or VLI induced
oscillations coexisting with standard oscillations.  Mixed scenarios of this
kind have  been considered, e.g., in \cite{Ga89,Co98,Ha96,Mi95} but, to our
knowledge, they have not been discussed on the basis of Super-Kamiokande
atmospheric $\nu$ observations so far.

In the presence of several concurrent processes leading to 
$\nu_\mu\leftrightarrow\nu_\tau$ oscillations, the global Hamiltonian $H$ is
the sum of several $2\times 2$ matrices $H_n$, which can be  diagonalized
through separate rotations (with angles $\xi_n$) of the flavor basis
$(\nu_\mu,\nu_\tau)$.  As far as models of the kind $\lambda^{-1}\propto E^{n}$
are concerned, the rotation angle $\xi$, which diagonalizes the total
hamiltonian, is related to the oscillation length $\lambda$ through equations
of the form \cite{Co98}
\begin{mathletters}
\begin{eqnarray}
\pi\lambda^{-1}\sin2\xi &=& \left| 
\sum_n c_n\, \sin 2\xi_n\, E^n\, e^{i\eta_n}\right|\ , 
\label{lambdacsia}\\
\pi\lambda^{-1}\cos2\xi &=& \sum_n c_n\, \cos 2\xi_n\, E^n\ ,
\label{lambdacsib}
\end{eqnarray}
\end{mathletters}
where the coefficients $c_n$ parametrize the strength of each  oscillation
mechanism. In general, only one of the complex phase factors $e^{i\eta_n}$ can
be rotated away, the others being physically observable \cite{Co98,Ha96}.  For
any given choice of the parameters $(c_n,\xi_n,\eta_n)$, one has to derive
the values of $\lambda$ and $\xi$ from the previous equations, and insert them
in Eq.~(\ref{P}) to get the flavor transition probability.

In the specific case of standard+VEP $(n=-1\oplus n=+1)$ oscillations,  
Eqs.~(\ref{lambdacsia}) and (\ref{lambdacsib}) can be rewritten as \cite{Ha96}
\begin{mathletters}
\begin{eqnarray}
\pi\,\lambda^{-1}\,\sin{2\xi}
&=& 
\left| 
1.27 \, \frac{\Delta m^2}{E}\, \sin2\psi
+
5.07\,  \frac{|\phi|\Delta\gamma}{10^{-21}}\,E \,
\sin{2\theta_G}\,e^{i\eta}
\right|\ , 
\label{pilambdaa}\\
\pi\,\lambda^{-1}\,\cos{2\xi}
&=& 
1.27 \, \frac{\Delta m^2}{E}\, \cos2\psi
+
5.07\,  \frac{|\phi|\Delta\gamma}{10^{-21}}\,E \,
\cos{2\theta_G}
\ , 
\label{pilambdab}
\end{eqnarray}
\label{pilambda}
\end{mathletters}
where the following units have been used: $[\Delta m^2]=10^{-3}$ eV$^2$, 
$[L]=[\lambda]=10^3$ km, and $[E]= {\rm GeV}$. The same equations formally
apply to violations of Lorentz invariance, modulo the replacements 
$|\phi|\Delta\gamma\to\delta v/2$ and $\theta_G\to\theta_v$ \cite{Gl97}. Notice
that the oscillation phase $\pi\lambda^{-1}L$, to be inserted in Eq.~(\ref{P}),
is proportional to the geometric average of the right hand sides of the above
equations, so it will contain, besides the standard term $(\propto E^{-1})$ and
the VEP term $(\propto E)$, also an energy-independent interference term. The
oscillation amplitude $\sin^2 2\xi$ also acquires a nontrivial dependence on
the neutrino energy through Eq.~(\ref{pilambda}).

\subsection{Constraints from Super-Kamiokande data}

Before performing a detailed fit to the SK data in the standard+VEP 
oscillation scenario,  let us derive some qualitative bounds on the magnitude
of possible VEP terms. According to the conclusions of Sec.~II, we expect that
the second (VEP) term on the right hand of Eqs.~(\ref{pilambdaa}) and
(\ref{pilambdab}) should be typically much smaller than the first (standard)
term, namely 
\begin{equation}
5.07\,\frac{|\phi|\Delta\gamma}{10^{-21}}\,\frac{E}{\rm GeV} \ll 1.27\,
\frac{\Delta m^2}{10^{-3}{\rm\ eV}^2}\,\frac{\rm GeV}{E}\ ,
\end{equation}
or equivalently (for $\Delta m^2\sim {\rm\ few}\times 10^{-3}$ eV$^2$):
\begin{equation}
|\phi|\Delta\gamma \ll 10^{-21} \left( \frac{\rm GeV}{E}\right)^2\ .
\label{e2}
\end{equation}
Therefore, the constraints to VEP effects should be stronger in the
highest-energy SK data sample (UP$\mu$ events). Since the parent neutrino
energy spectrum for UP$\mu$'s  is peaked around $10^2$ GeV,  the sensitivity to
VEP-induced oscillations is expected to reach levels of $O(10^{-25})$ in the
parameter $|\phi|\Delta\gamma$  (or, equivalently, in the parameter $\delta
v/2$ for the VLI case). Of course, such sensitivity depends somewhat on 
$\theta_G$.

In order to get some insight about the  $\theta_G$-dependence of the expected
constraints, let us consider the extreme values $\theta_G=0$ and
$\theta_G=\pi/4$, and take $(\Delta m^2,\sin^2 2\psi)$ at their best-fit values
$(2.8\times 10^{-3}{\rm\ eV}^2,1)$. We also fix $e^{i\eta}=1$ for simplicity.
Then $P_{\mu\tau}$ takes a simple form,
\begin{equation}
P_{\mu\tau}=\left\{
\begin{array}{ll}
\displaystyle\frac{1}{1+x^2}\sin^2\left(A\sqrt{1+x^2}\right) &,\ \theta_G=0\ ,
 \\ \\
\sin^2\Big(A(1+x)\Big) &,\ \theta_G=\pi/4\ ,
\end{array}
\right.
\label{px}
\end{equation}
where $A=3.56\,L/E$, $x=1.42\,E^2|\phi|\Delta\gamma/10^{-21}$, and the units
are $[E]={\rm GeV}$ and $[L]=10^3{\rm\ km}$. The standard case   ($P^{\rm
std}_{\mu\tau}=\sin^2A$) is recovered for $x=0$. In the UP$\mu$ event sample,
where the VEP effect is larger, the value of $A$ is typically small $(\lesssim
0.6)$, and one can easily check numerically  [from Eq.~(\ref{px})] that the
difference $P_{\mu\tau}-P^{\rm std}_{\mu\tau}$ grows more rapidly with $x$  for
$\theta_G=\pi/4$ than for $\theta_G=0$. Therefore, we expect a higher
sensitivity to  $x$ (i.e., to $|\phi|\Delta\gamma$) at larger $\theta_G$. 
Moreover, at small values of both $A$ and $x$ it turns out that
$P_{\mu\tau}-P^{\rm std}_{\mu\tau}<0$ ($>0$) for $\theta_G=0$  
($\theta_G=\pi/4$), so that the muon rates should be less (more) suppressed
than in the standard oscillation case.

Figure~5 illustrates quantitatively the previous considerations, by showing the
VEP effect (added to best-fit standard oscillations) on the SK muon
distributions for two  representative cases: (i) $\theta_G=0$ and
$|\phi|\Delta\gamma=1.5\times 10^{-24}$  (solid line, blue in color); (ii)
$\theta_G=\pi/4$ and  $|\phi|\Delta\gamma=2.0\times 10^{-26}$  (dotted line,
green in color). We anticipate that the such values are close to the border of
the parameter region excluded by SK. The standard oscillation curves
($|\phi|\Delta\gamma=0$) are also shown for reference (dashed lines, red in
color); they are identical to the best fit curves  for standard oscillations
($n=-1$) in Fig.~2.  As expected from the preceding discussion, the VEP effect
is manifest at high energies (UP$\mu$ sample), and the expected deficit of
UP$\mu$'s is more pronounced for $\theta_G=\pi/4$ than for $\theta_G=0$.

The next step is to perform a $\chi^2$ analysis of the standard+VEP scenario.
We take for the moment $(\Delta m^2,\sin^2 2\theta)=(2.8\times 10^{-3}{\rm\
eV}^2,\,1)$ and $e^{i\eta}=1$, while leaving the parameters
($|\phi|\Delta\gamma, \theta_G$) free.  We find an important result, that
strengthens the conclusions of Sec.~II: The  $\chi^2$ fit is never improved in
the presence of VEP effects, as compared with the value $\chi^2_{\min}=20.3$
derived in Fig.~2 for the standard  case. Thus, not only VEP-induced
oscillations cannot be the {\em leading\/} mechanism underlying the SK
observations, but also there is no indication in favor of {\em subleading\/}
VEP oscillation terms. As a  consequence, we can place
well-defined upper bounds on  violations of relativity principles in the
$(\nu_\mu,\nu_\tau)$ sector.

Figure~6 shows the 90\% and 99\% C.L. limits on $|\phi|\Delta\gamma$ as a
function of the VEP mixing parameter $\sin^22\theta_G$. The same limits apply
to the neutrino asymptotic speed difference $\delta v/2$, as a function of the
VLI mixing parameter $\sin^22\theta_v$.  As expected from the discussion at the
beginning of this subsection,  the limits obtained in Fig.~6 are roughly of
$O(10^{-25})$, and become stronger as $\theta_G$ increases. The stringent
bounds in Fig.~6, together with the results shown in Fig.~3, represent our main
contribution to the current understanding of atmospheric neutrino oscillations
induced by violations of relativity principles in the $(\nu_\mu,\nu_\tau)$
sector. Notice that such bounds pre-empt the region of VEP (or VLI) parameters
explorable with proposed long-baseline accelerator neutrino facilities 
\cite{Ha96}.%
\footnote{Notice that the values of $L$, $E$, and $L\cdot E$ probed in the
UP$\mu$ sample by Super-Kamiokande are higher than in proposed long-baseline
neutrino beams.}

Finally, we have investigated the robustness of the bounds shown in Fig.~6
under variations of the standard mass-mixing parameters. We have repeated the
fit by varying $\Delta m^2$ and $\sin^2 2\psi$ within the 90\% C.L. limits
shown in Fig.~4, and also by taking negative values for $\Delta\gamma$, as well
as generic values  for the complex phase $\eta$. We have found that, in any
case, the value of $\chi^2_{\min}$ is not smaller than in the best-fit standard
case. Therefore, standard+VEP (or standard+VLI) oscillations never represent a
better description of the SK data, as compared to best-fit standard
oscillations.  Under the above variations,  the upper bounds shown in Fig.~6
are somewhat modified within factors of a few, but do not change qualitatively:
they always become stronger as $\sin^2 2\theta_G$ increases. The most
conservative upper bound (including negative $\Delta \gamma$ cases) turns out
to be
\begin{equation}
|\phi\Delta\gamma| < 3 \times 10^{-24} {\rm\ \  at\ 90\%\ C.L.}\ ,
\end{equation}
independently of $\theta_G$. In the specific case $\theta_G=\pi/4$,  the above
bound can be lowered at least to $\lesssim 10^{-25}$. 
Analogous limits apply to the VLI parameters $|\delta v|/2$ and $\theta_v$. 
In particular,
\begin{equation}
|\delta v| < 6 \times 10^{-24} {\rm\ \  at\ 90\%\ C.L.}\ 
\end{equation}

To our knowledge, the above limits to violations of special or general
relativity principles are the strongest ever placed in the $(\nu_\mu,\nu_\tau)$
sector. They are valid under an assumption which is supported by the present
data and appears likely to be corroborated in the future, namely, that standard
$\nu_\mu\leftrightarrow\nu_\tau$ oscillations generated by $\nu$ mass and
mixing represent the dominant mechanism underlying the Super-Kamiokande
observations.

\section{Summary and Conclusions}

Among the $\nu_\mu\leftrightarrow\nu_\tau$ models with oscillation length
following a power-law energy dependence, standard neutrino oscillations
generated by $\nu$ mass and mixing are unique in providing a good description
of the Super-Kamiokande atmospheric neutrino data, and are strongly favored as
leading mechanism for $(\nu_\mu,\nu_\tau)$ flavor transitions. Additional,
subleading $\nu_\mu\leftrightarrow\nu_\tau$ oscillations generated by possible
violations of special or general relativity in the neutrino sector do not
improve the agreement with the data, and must thus have a relatively small (or
zero) amplitude. In particular, the fractional difference of asymptotic $\nu$
velocities $|\delta v|/2$ or of $\nu$ couplings to gravity $|\phi\Delta\gamma|$
cannot exceed the value $\sim 3\times 10^{-24}$ at 90\% C.L.\ for unconstrained
neutrino mixing.  The broadness of the neutrino energy range probed by
Super-Kamiokande is crucial to obtain  such strong limits.

\acknowledgements

We thank M.\ Gasperini for inspiring discussions and for useful comments. The
work of A.M.\ and G.S.\ is supported by the Italian Ministero
dell'Universit\`a  e della Ricerca Scientifica e Tecnologica through a PhD
grant.


\newcommand{\InsertFigure}[2]{\newpage\begin{center}\mbox{%
\epsfig{bbllx=1.4truecm,bblly=1.3truecm,bburx=19.5truecm,bbury=26.5truecm,%
height=21.2truecm,figure=#1}}\end{center}\vspace*{-1.6truecm}%
\parbox[t]{\hsize}{\small\baselineskip=0.45truecm\hskip0.5truecm #2}}

\InsertFigure{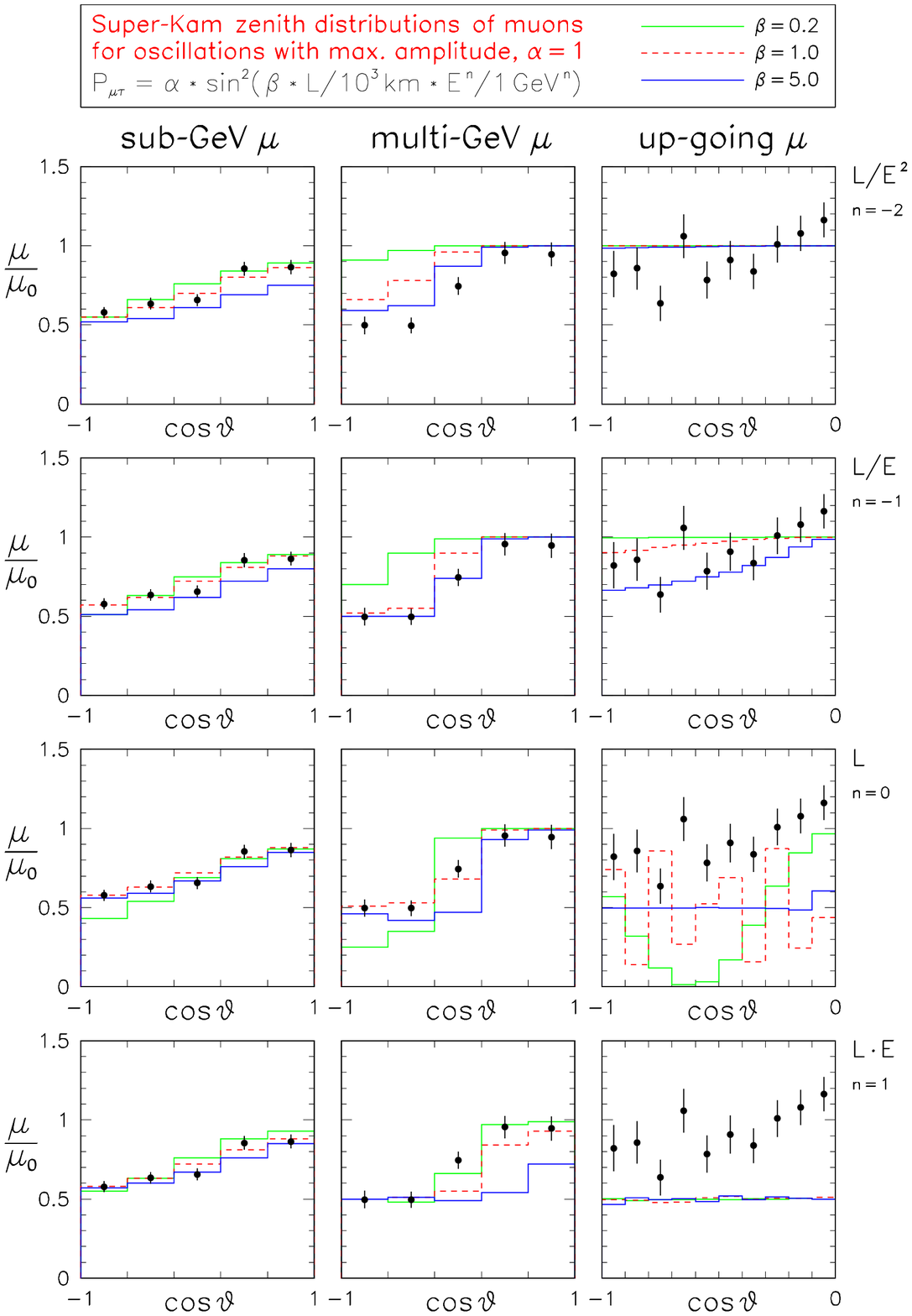}%
{FIG.~1.   Expected zenith distributions of SG$\mu$, MG$\mu$,  and UP$\mu$
events in Super-Kamiokande for maximal mixing  ($\alpha=1$) and for
representative values of $\beta$: dotted line, $\beta=0.2$; dashed line,
$\beta=1$;  solid line, $\beta=5$ (on color printers such lines appear in
green, red, and blue, respectively). In each bin, the muon rates $\mu$  are
normalized to the expectations $\mu_0$ in the absence of oscillations. The 45
kTy SK data  (dots with statistical error bars) are superposed to guide the
eye. From top to bottom, the energy exponent $n$ takes the values $n=-2$, $-1$,
0, and 1.}
\InsertFigure{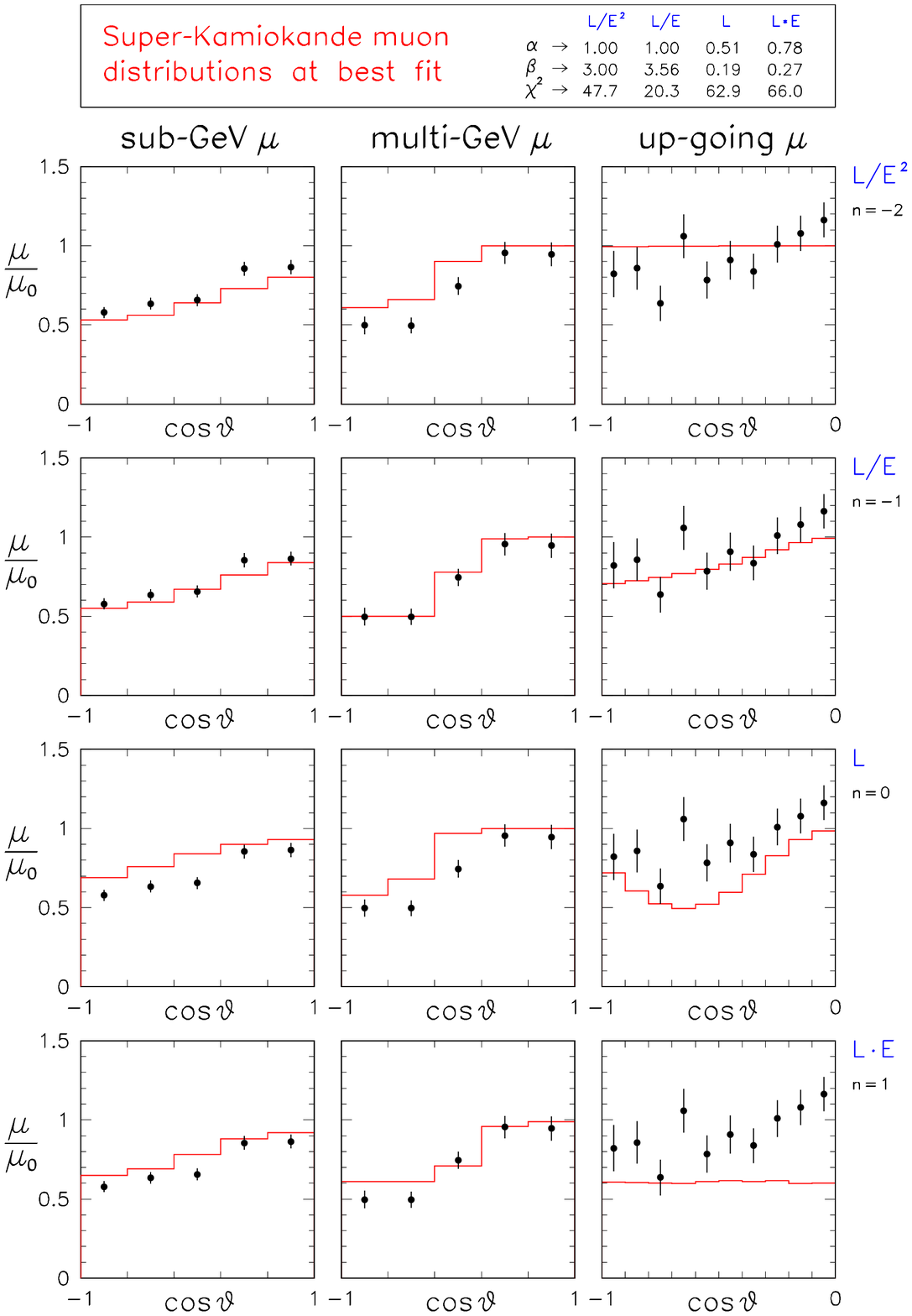}%
{FIG.~2.  Best-fits to the zenith distributions of muons in SK for the four
cases considered in Fig.~1 ($n=-2$, $-1$, 0, and 1), as obtained through a
$\chi^2$ analysis of all the data ($\mu$-like and $e$-like events) with
unconstrained $\alpha$ and $\beta$. The values of $\alpha$, $\beta$, and
$\chi^2$ at the best-fit points are  reported in the top part of the  figure.}
\InsertFigure{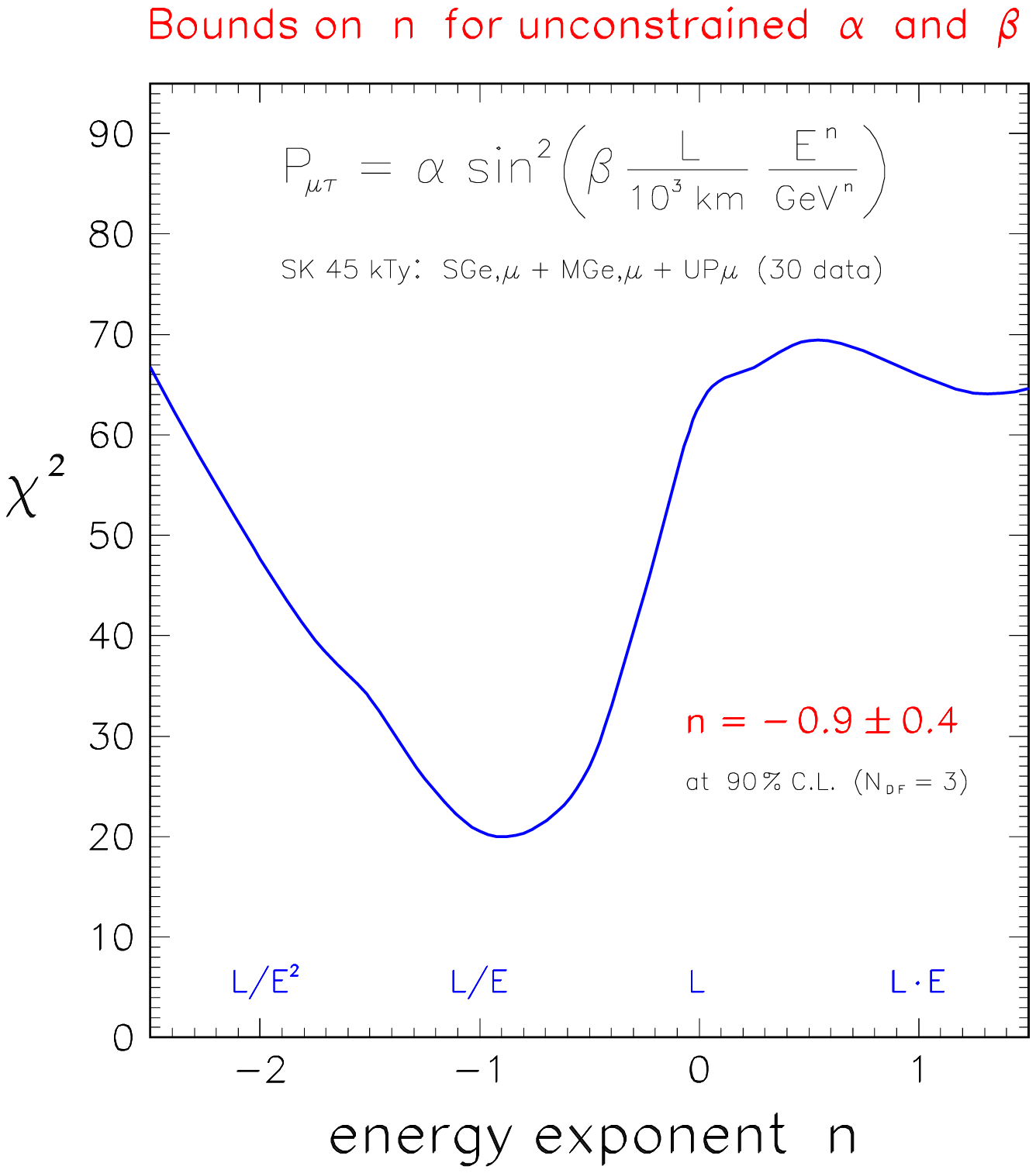}%
{FIG.~3.  $\chi^2$ function from the  fit to the SK data, assuming continuous
values of the energy exponent $n$ and unconstrained scale factors for the
oscillation amplitude $\alpha$ and phase $\beta$. The standard case
($n=-1$) is very close to the best fit.}
\InsertFigure{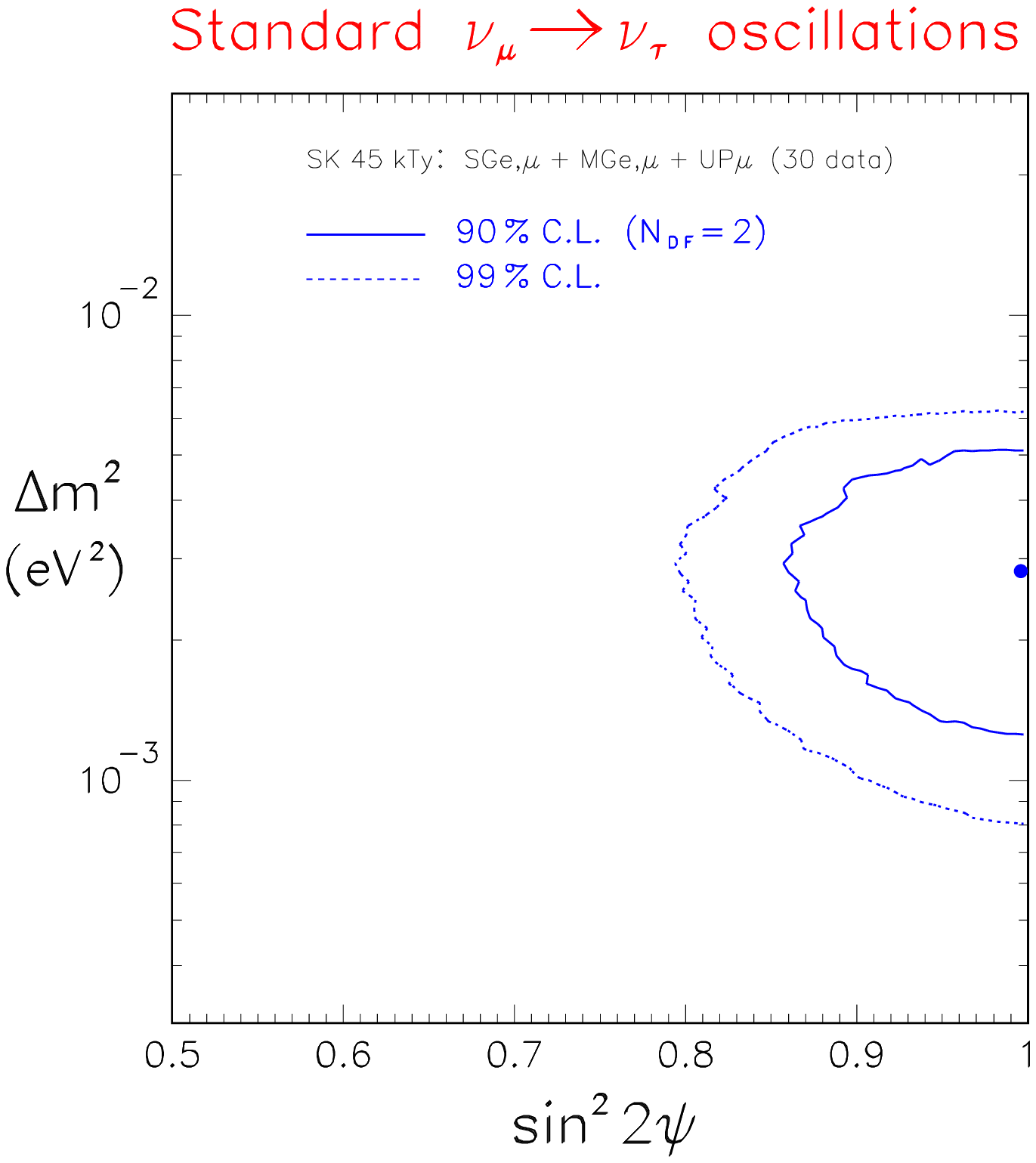}%
{FIG.~4.  Updated bounds on the neutrino mass-mixing parameters for standard
$\nu_\mu\leftrightarrow\nu_\tau$ oscillations, as derived by our global
analysis of all the Super-Kamiokande atmospheric neutrino data.}
\InsertFigure{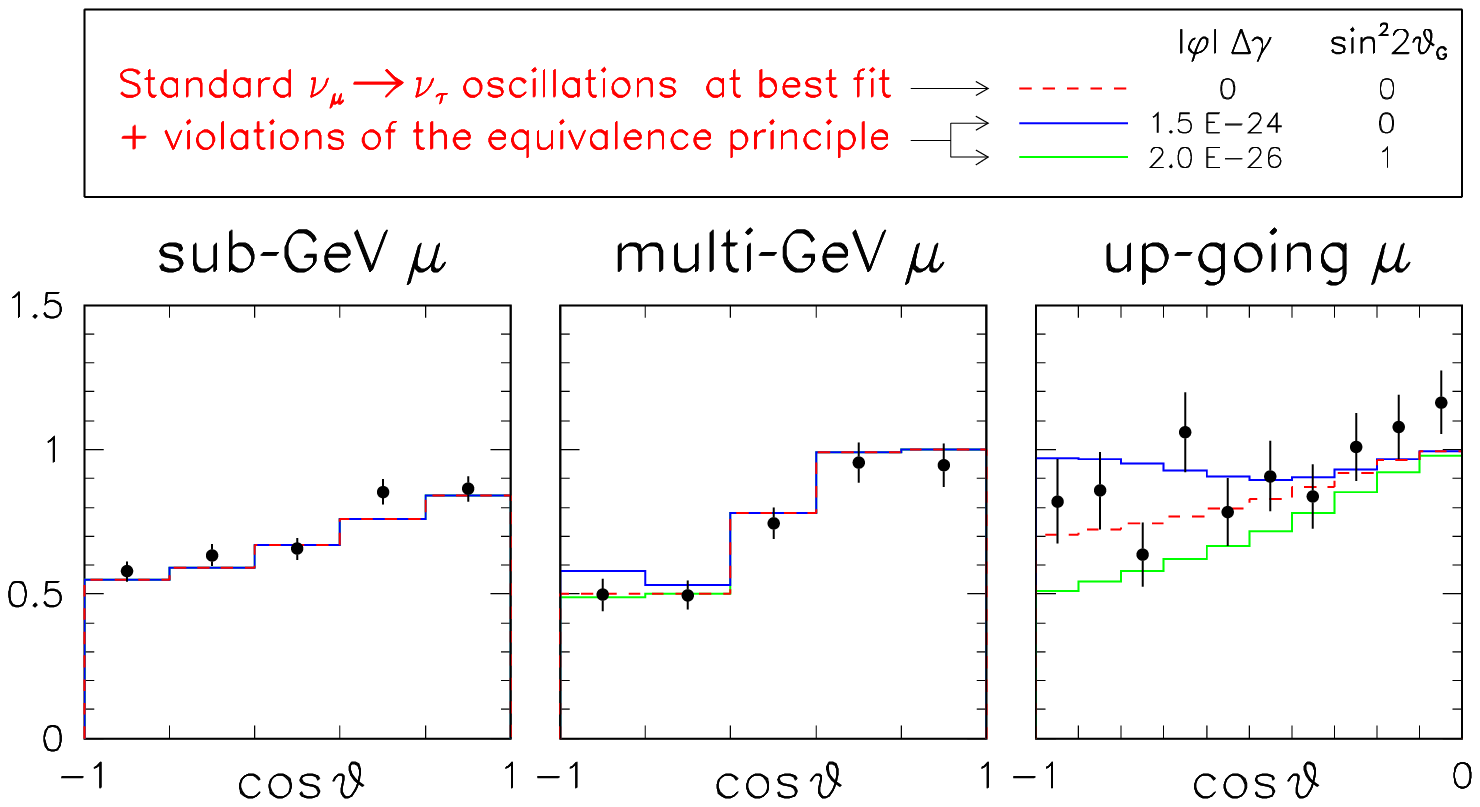}%
{FIG.~5.  Effect of subleading oscillations induced by violations of the
equivalence principle (VEP) on the muon distributions in Super-Kamiokande.
Standard mass-mixing oscillation parameters are taken at their best fit, and
the complex phase $e^{i\eta}$ is taken equal to 1. Dashed line (red in color):
no VEP. Solid line (blue in color): VEP with  $|\phi|\Delta\gamma=1.5\times
10^{-24}$ and $\theta_G=0$. Dotted line (green in color): VEP with 
$|\phi|\Delta\gamma=2\times 10^{-26}$ and $\theta_G=\pi/4$.}
\InsertFigure{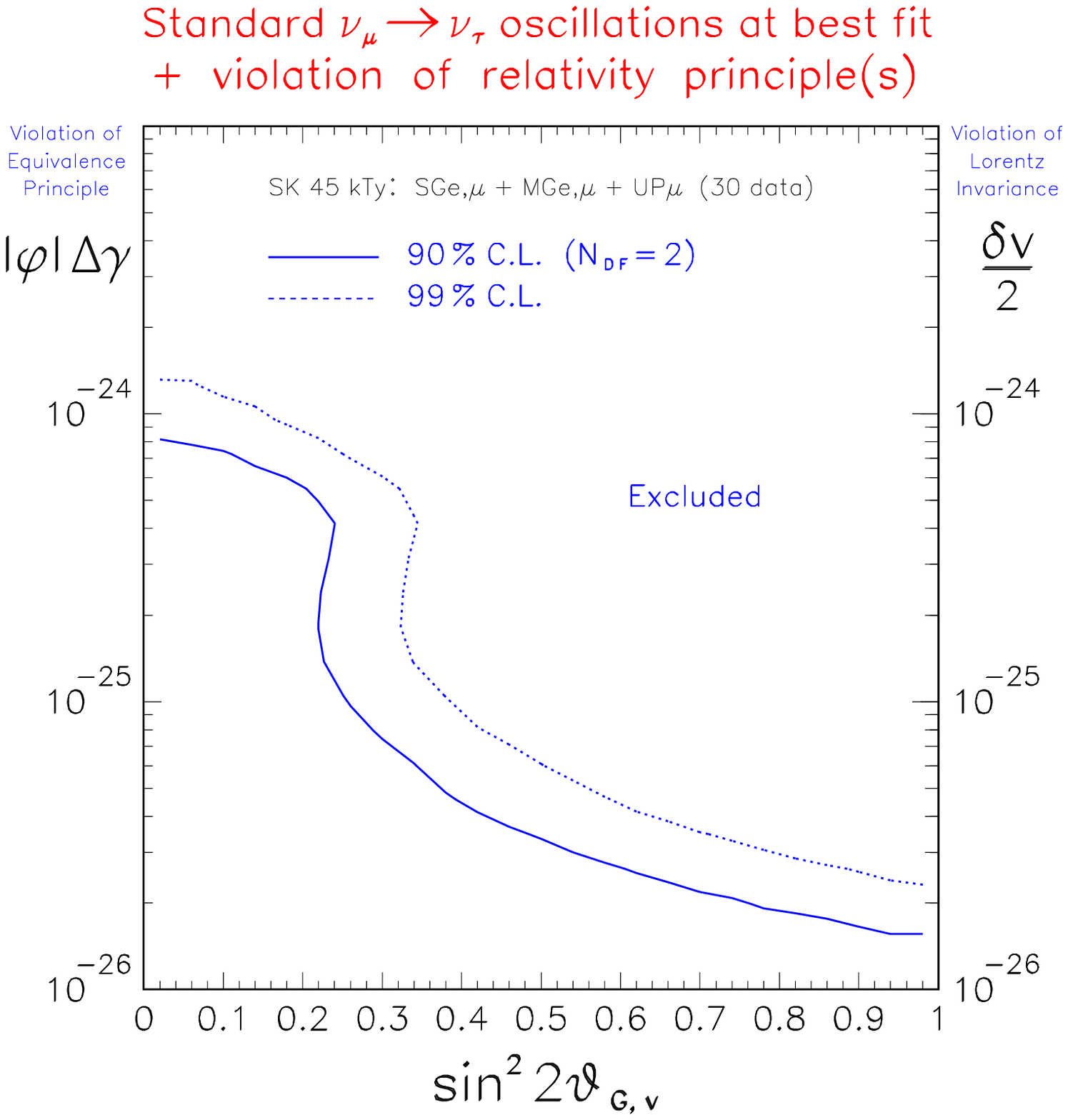}%
{FIG.~6. Bounds on the parameters that characterize violations of special or
general relativity principles, assumed to generate subleading
$\nu_\mu\leftrightarrow\nu_\tau$ flavor transitions concurrent with standard
(leading) neutrino oscillations. Best-fit neutrino mass-mixing values are
assumed. See the text for further details.}


\end{document}